\documentclass[mypaper,7pt,twoside]{CoAst}
\usepackage{epsf,graphicx,fancyhdr,natbib,amssymb}
\renewcommand{\chaptermark}[1]{\markboth{#1}{}}

\renewcommand{\headrulewidth}{0pt}
\newcommand{\apj}{ApJ}
\newcommand{\aap}{A\&A}  
\newcommand{\mnras}{MNRAS}

\newcommand{\teff}{\ensuremath{T_{eff}}}             

\newcommand{\kopf}{\small\itshape Comm. in Asteroseismology\\ Vol. number, publication date (will be inserted in the production process)}
\newcommand{\Authors}[1]{\begin{center}\normalsize\bf\sf #1 \end{center}}

\renewcommand{\author}[1]{\begin{center}\normalsize\bf\sf #1 \end{center}}
\newcommand{\Address}[1]{\begin{center}\small\sf #1 \end{center}}

\renewenvironment{abstract}{\section*{Abstract}\normalsize\sf}{}
\newcommand{\References}[1]{\begin{flushleft}{\large References\\}\vspace*{2mm}\small #1 \end{flushleft}}

\newcommand{\chapterDSSN}[2]{\chapter[\sf\normalsize #1\\ \footnotesize \hspace*{5mm}by #2 \sf\normalsize][]{#1\\}\rhead[\fancyplain{}{\sf\footnotesize \center{#1}}]{\fancyplain{}{\sffamily\thepage}}\lhead[\fancyplain{\kopf}{\sffamily\thepage}]{\fancyplain{\kopf}{\sf\footnotesize \center{#2}}}}

\newcommand{\figureDSSN}[5]{\begin{figure}[#4]
\centering
\includegraphics*[#5]{#1}
\caption{#2}
\label{#3}
\end{figure}}

\renewcommand{\chaptername}{}
\newcommand{\acknowledgments}[1]{\vspace*{5mm}\noindent\begin{bf}Acknowledgments. \end{bf} #1}

\def\apss{Ap\&SS}
\def\apjl{ApJ}

\begin{document}
\sf

\chapterDSSN{Revised instability domains of SPB and $\beta$ Cephei stars}
{A. Miglio, J. Montalb\'an and  M.-A. Dupret}

\Authors{A. Miglio$^1$, J. Montalb\'an$^1$ and M.-A. Dupret$^2$}
\Address{
$^{1}$Institut d'Astrophysique, All\'ee du 6 Ao\^ut, 17, B-4000 Li\`ege, Belgium\\
$^{2}$Observatoire de Paris, LESIA, CNRS UMR~8109, 92195 Meudon, France
}

\noindent
\begin{abstract}
The excitation of pulsation modes in $\beta$ Cephei and Slowly Pulsating B stars is known
to be very sensitive to opacity changes in the stellar interior where $T\sim2\times 10^5\,\rm K$.
 In this region differences in opacity up to $\sim 50\%$ can be induced by the choice between
 OPAL and OP opacity tables, and between two different metal mixtures (\citealt{Grevesse93} and
 \citealt{Asplund05}). We have extended the non-adiabatic computations presented in \citet{Miglio07}
 towards models of higher mass and pulsation modes of degree $\ell=3$, and
 we  present here the instability domains in the HR- and $\log{P}$-$\log{T_{\rm eff}}$
 diagrams resulting from different choices of opacity tables, and for three different
 metallicities.
\end{abstract}

\section{Introduction}
The detection of B-type pulsators in low metallicity environments (see e.g. \citealt{Kolaczkowski06}
and references therein), and the large number of pulsation modes detected in B stars, are now revealing
new discrepancies between theory and observations that challenge standard stellar models.
For instance, the two $\beta$~Cep stars 12 Lacertae  and $\nu$~Eridani present low order p-modes
with frequencies higher than those predicted by pulsation models,  as well as high-order g-modes
(SPB type oscillation) \citep[][ and references therein]{Jerzykiewicz05,Handler06}.

The interpretation of observations in the framework of standard stellar models must
take into consideration the uncertainties in the basic input physics.
In fact, pulsations modes in SPBs and $\beta$~Cep stars are excited by the
$\kappa$-mechanism \citep[see e.g.][]{Dziembowski93} due to the Fe-group opacity bump
at $T \sim 2\times 10^5$~K, and in the last three years there have been two important updates
of the basic physics that can affect the study of B-type pulsators:
{\it i)} the revised solar metal mixture \citep{Asplund05} that implies a 25\% larger Fe mass fraction for
a given metallicity $Z$; and {\it ii)} the new Fe data included in OP opacity computations that lead to
an opacity in the Z-bump increased by 18\% with respect to the previous values \citep{Badnell05}.

\citet{Miglio07} (hereafter Paper I) and \citet{Pamyatnykh07} showed that the combination these updates have a remarkable effect  on the instability domains of SPB and $\beta$~Cephei pulsators compared to the results obtained using OPAL opacities \citep{Iglesias96} and \citet{Grevesse93}
metal mixture.

In Paper~I we analyzed the role of chemical composition and opacity computations on
the instability strip of B-type pulsators and on the frequency domain of expected
excited modes. In the present paper  we have extended the computations presented in Paper~I by
considering stellar masses up to 18~$\rm M_\odot$ instead of 12~$\rm M_\odot$, and by
carrying out the non-adiabatic analysis also for $\ell=3$ modes.
Only throughout comparisons with observations we will be able to assess if, and to which extent,
the current uncertainties on opacity calculations and on the assumed metal mixture are able
 to explain the discrepancies between recent observations and standard stellar models.
For this purpose we present in the following sections the instability strips in the
HR  ($\log{L}$-$\log{T_{eff}}$) and in the period-effective temperature ($\log{P}$-$\log{T_{eff}}$)
diagrams resulting from the non-adiabatic calculations presented in Paper I and
extended as mentioned above.

\section{Stellar models and opacities}
We computed stellar models with the code CLES (Code Li\'egeois d'Evolution Stellaire, \citealt{Scuflaire07}).
The main physical inputs are:
OPAL2001 equation of state \citep{Rogers02} and  \cite{Caughlan88} nuclear reaction rates  with \cite{Formicola04}
for the $^{14}$N(p,$\gamma$)$^{15}$O cross-section. Convective transport is treated by using the
classical Mixing Length Theory of convection \citep{Bohm58}, and a convective overshooting
parameter of 0.2 pressure scale height was assumed in all the models.
For the chemical composition we have considered: \citet{Grevesse93} (GN93) and \citet{Asplund05} corrected with the Ne abundance determined by \citet{Cunha06} (AGS05+Ne).
We have computed models with: {\it i)} OPAL opacity tables with GN93 and {\it ii)} AGS05+Ne chemical composition, then models with {\it iii)} OP opacity tables assuming GN93 and {\it iv)} AGS05+Ne mixtures.
All the opacity tables are completed at $\log T < 4.1$  with the corresponding GN93 and AGS05 low temperature tables by \citet{Ferguson05}.

The masses considered span from 2.5 to 18~$\rm M_\odot$, and the chemical compositions considered are:
$X=0.70$ for the hydrogen mass fraction, and three different metal mass fractions: $Z=0.02$, 0.01 and 0.005. For all the models the evolution was followed from the Pre-Main Sequence.

We recall that the differences between opacities computed with OPAL, OP and with the metal mixtures
considered can reach nearly 50\% in the region where the driving of pulsations occurs
($T\sim 2\times 10^5\,\rm K$) for a typical $\beta$ Cep star (see Paper I for a detailed comparison).
Though not included in the calculations presented here, it is worth recalling that the effect of
considering the \citet{Asplund05} metal mixture without the higher Neon abundance proposed by
\citet{Cunha06} is to further increase the Fe relative mass fraction by $\sim 5\%$.
In a 10~$\rm M_\odot$ model (and for a given value of $Z$) this induces a further increase of the
opacity at $T\sim 2\times 10^5\,\rm K$ up to $7\%$, that only slightly modifies the instability
strips presented here.

\section{Results: Updated SPBs and $\beta$-Cep instability domains}

We carry out a pulsational stability analysis of  main-sequence models from our grid using the
non-adiabatic code MAD \citep{Dupret03}. As mentioned above, in these computations we fixed the
overshooting parameter $\alpha_{\rm ov}$ at 0.2 and the initial hydrogen mass fraction $X$ at 0.70.
 For discussion about the effect of assuming different
$\alpha_{\rm ov}$ or $X$ on the stability domain, as well as for the stability study in post-MS models, we
refer to the work by \citet{Pamyatnykh99}.
We checked the stability of radial modes and of non-radial p- and g-modes of degree $1\le \ell \le 3$.

The location of the instability strip in the HR diagram and the frequency of the excited modes
are determined by the properties of the metal opacity bump.
The effects of the choice of the metal mixture (GN93 or AGS05+Ne) and of the opacity computations (OPAL or
OP) on the HR location of instability domains are shown in figures \ref{fig:is_z2}, \ref{fig:is_z1},
 and \ref{fig:is_z05}  for models
with metallicity $Z=$0.02, 0.01, and 0.005, respectively. The combined effects on the excited modes of OP
opacity and AGS05+Ne metal mixture, compared with the standard OPAL with GN93,
are also shown by means of the  Period-$T_{\rm eff}$ diagram in Fig.~\ref{fig:logP}.

The results presented in these figures can be summarized as follows:
\begin{enumerate}
\item Since the region where $\kappa_{\rm T}=\left(\partial\log{\kappa_{\rm R}}/\partial \log{T}\right)_\rho$
 increases outwards is found deeper in the star, with respect to
the models computed with  OPAL tables, the blue borders of the instability strips are hotter
with OP models compared to OPAL ones.
\item The $T_{\rm eff}$ domain for which we find SPB pulsators using OP opacities is $\sim 3000$~K larger than for OPAL models.
As a consequence, the number of expected hybrid $\beta$~Cep--SPB objects is also larger for OP models.
\item The impact  of the different OP--OPAL opacities is more important for low metallicity.
As shown in Fig.~2, while OPAL--GN93 models with Z=0.01 are hardly able to produce
a narrow instability strip at the end of MS, with excited modes only for $\ell>1$,
the OP models present $\ell=$0--3 excited modes already for an evolutionary state corresponding to $X_c \simeq 0.3$.
\item  The Fe-mass fraction enhancement in the AGS05+Ne mixture, compared with GN93,
has the main effect of extending towards higher overtones the
range of excited frequencies.
\item Furthermore,  while the different profile of $\kappa$ in OP and OPAL computations modifies
the blue border of the instability strip, a larger Fe-mass fraction in the metal mixture
provides a slightly wider instability bands, and this effect increases as the metallicity decreases.
Thus, the number of $\beta$~Cep pulsators expected with AGS05+Ne is more than three times larger
than with GN93.
\item   Computations for the lowest metallicity considered (Z=0.005), show that none
of the different OP/OPAL and GN93/AGS05+Ne
evolutionary tracks for masses up to 18~$M_{\odot}$ predicts $\beta$~Cep pulsators,
whereas we find SPB-type modes excited when considering OP with AGS05+Ne.
\end{enumerate}


The instability strips presented in this work will be made available to the community via the HELAS (European
Helio- and Asteroseismology Network) website\footnote{\texttt{http://http://www.helas-eu.org}} and are also available upon request to the authors.

\acknowledgments{The authors are thankful to R. Scuflaire for his kind help with CLES. A.M. and J.M. acknowledge financial support from the Prodex-ESA Contract Prodex 8 COROT (C90199).

}

\References{
\bibitem[{{Asplund} {et~al.}(2005){Asplund}, {Grevesse}, {Sauval}, {Allende
  Prieto}, \& {Blomme}}]{Asplund05}
{Asplund}, M., {Grevesse}, N., {Sauval}, A.~J., {Allende Prieto}, C., \&
  {Blomme}, R. 2005, \aap, 431, 693

\bibitem[{{B{\" o}hm-Vitense}(1958)}]{Bohm58}
{B{\" o}hm-Vitense}, E. 1958, Zeitschrift fur Astrophysics, 46, 108

\bibitem[{{Badnell} {et~al.}(2005){Badnell}, {Bautista}, {Butler}, {Delahaye},
  {Mendoza}, {Palmeri}, {Zeippen}, \& {Seaton}}]{Badnell05}
{Badnell}, N.~R., {Bautista}, M.~A., {Butler}, K., {et~al.} 2005, \mnras, 360,
  458

%
\bibitem[{{Caughlan} \& {Fowler}(1988)}]{Caughlan88}
{Caughlan}, G.~R. \& {Fowler}, W.~A. 1988, Atomic Data and Nuclear Data Tables,
  40, 283

\bibitem[{{Cunha} {et~al.}(2006){Cunha}, {Hubeny}, \& {Lanz}}]{Cunha06}
{Cunha}, K., {Hubeny}, I., \& {Lanz}, T. 2006, \apjl, 647, L143

\bibitem[{{Dupret} {et~al.}(2003){Dupret}, {De Ridder}, {De Cat}, {Aerts},
  {Scuflaire}, {Noels}, \& {Thoul}}]{Dupret03}
{Dupret}, M.-A., {De Ridder}, J., {De Cat}, P., {et~al.} 2003, \aap, 398, 677

\bibitem[{{Dziembowski} {et~al.}(1993){Dziembowski}, {Moskalik}, \&
  {Pamyatnykh}}]{Dziembowski93}
{Dziembowski}, W.~A., {Moskalik}, P., \& {Pamyatnykh}, A.~A. 1993, \mnras, 265,
  588

\bibitem[{{Ferguson} {et~al.}(2005){Ferguson}, {Alexander}, {Allard}, {Barman},
  {Bodnarik}, {Hauschildt}, {Heffner-Wong}, \& {Tamanai}}]{Ferguson05}
{Ferguson}, J.~W., {Alexander}, D.~R., {Allard}, F., {et~al.} 2005, \apj, 623,
  585

\bibitem[{{Formicola} {et~al.}(2004){Formicola}, {Imbriani}, {Costantini},
  {Angulo}, {Bemmerer}, {Bonetti}, {Broggini}, {Corvisiero}, {Cruz},
  {Descouvemont}, {F{\"u}l{\"o}p}, {Gervino}, {Guglielmetti}, {Gustavino},
  {Gy{\"u}rky}, {Jesus}, {Junker}, {Lemut}, {Menegazzo}, {Prati}, {Roca},
  {Rolfs}, {Romano}, {Rossi Alvarez}, {Sch{\"u}mann}, {Somorjai}, {Straniero},
  {Strieder}, {Terrasi}, {Trautvetter}, {Vomiero}, \&
  {Zavatarelli}}]{Formicola04}
{Formicola}, A., {Imbriani}, G., {Costantini}, H., {et~al.} 2004, Physics
  Letters B, 591, 61

\bibitem[{{Grevesse} \& {Noels}(1993)}]{Grevesse93}
{Grevesse}, N. \& {Noels}, A. 1993, in La formation des \'el\'ements chimiques,
  AVCP, ed. R.~D. Hauck~B., Paltani~S., 205--257

\bibitem[{{Handler} {et~al.}(2006){Handler}, {Jerzykiewicz},
  {Rodr{\'{\i}}guez}, {Uytterhoeven}, {Amado}, {Dorokhova}, {Dorokhov},
  {Poretti}, {Sareyan}, {Parrao}, {Lorenz}, {Zsuffa}, {Drummond},
  {Daszy{\'n}ska-Daszkiewicz}, {Verhoelst}, {De Ridder}, {Acke}, {Bourge},
  {Movchan}, {Garrido}, {Papar{\'o}}, {Sahin}, {Antoci}, {Udovichenko},
  {Csorba}, {Crowe}, {Berkey}, {Stewart}, {Terry}, {Mkrtichian}, \&
  {Aerts}}]{Handler06}
{Handler}, G., {Jerzykiewicz}, M., {Rodr{\'{\i}}guez}, E., {et~al.} 2006,
  \mnras, 365, 327

\bibitem[{{Iglesias} \& {Rogers}(1996)}]{Iglesias96}
{Iglesias}, C.~A. \& {Rogers}, F.~J. 1996, \apj, 464, 943

\bibitem[{{Jerzykiewicz} {et~al.}(2005){Jerzykiewicz}, {Handler}, {Shobbrook},
  {Pigulski}, {Medupe}, {Mokgwetsi}, {Tlhagwane}, \&
  {Rodr{\'{\i}}guez}}]{Jerzykiewicz05}
{Jerzykiewicz}, M., {Handler}, G., {Shobbrook}, R.~R., {et~al.} 2005, \mnras,
  360, 619

\bibitem[{{Ko{\l}aczkowski} {et~al.}(2006){Ko{\l}aczkowski}, {Pigulski},
  {Soszy{\'n}ski}, {Udalski}, {Kubiak}, {Szyma{\'n}ski}, {{\.Z}ebru{\'n}},
  {Pietrzy{\'n}ski}, {Wo{\'z}niak}, {Szewczyk}, \&
  {Wyrzykowski}}]{Kolaczkowski06}
{Ko{\l}aczkowski}, Z., {Pigulski}, A., {Soszy{\'n}ski}, I., {et~al.} 2006,
  Memorie della Societa Astronomica Italiana, 77, 336

\bibitem[{{Miglio} {et~al.}(2007){Miglio}, {Montalb{\'a}n}, \&
  {Dupret}}]{Miglio07}
{Miglio}, A., {Montalb{\'a}n}, J., \& {Dupret}, M.-A. 2007, \mnras, 375, L21

\bibitem[{{Pamyatnykh}(1999)}]{Pamyatnykh99}
{Pamyatnykh}, A.~A. 1999, Acta Astronomica, 49, 119

\bibitem[{{Pamyatnykh} \& {Ziomek}(2007)}]{Pamyatnykh07}
{Pamyatnykh}, A.~A. \& {Ziomek}, W. 2007, CoAst, in press

\bibitem[{{Rogers} \& {Nayfonov}(2002)}]{Rogers02}
{Rogers}, F.~J. \& {Nayfonov}, A. 2002, \apj, 576, 1064

\bibitem[{{Scuflaire} {et~al.}(2007){Scuflaire}, {Th{\'e}ado}, {Montalb{\'a}n},
  {Miglio}, {Bourge}, {Godart}, {Thoul}, \& {Noels}}]{Scuflaire07}
{Scuflaire}, R., {Th{\'e}ado}, S., {Montalb{\'a}n}, J., {et~al.} 2007, \apss,
  in press

}
\newpage
\figureDSSN{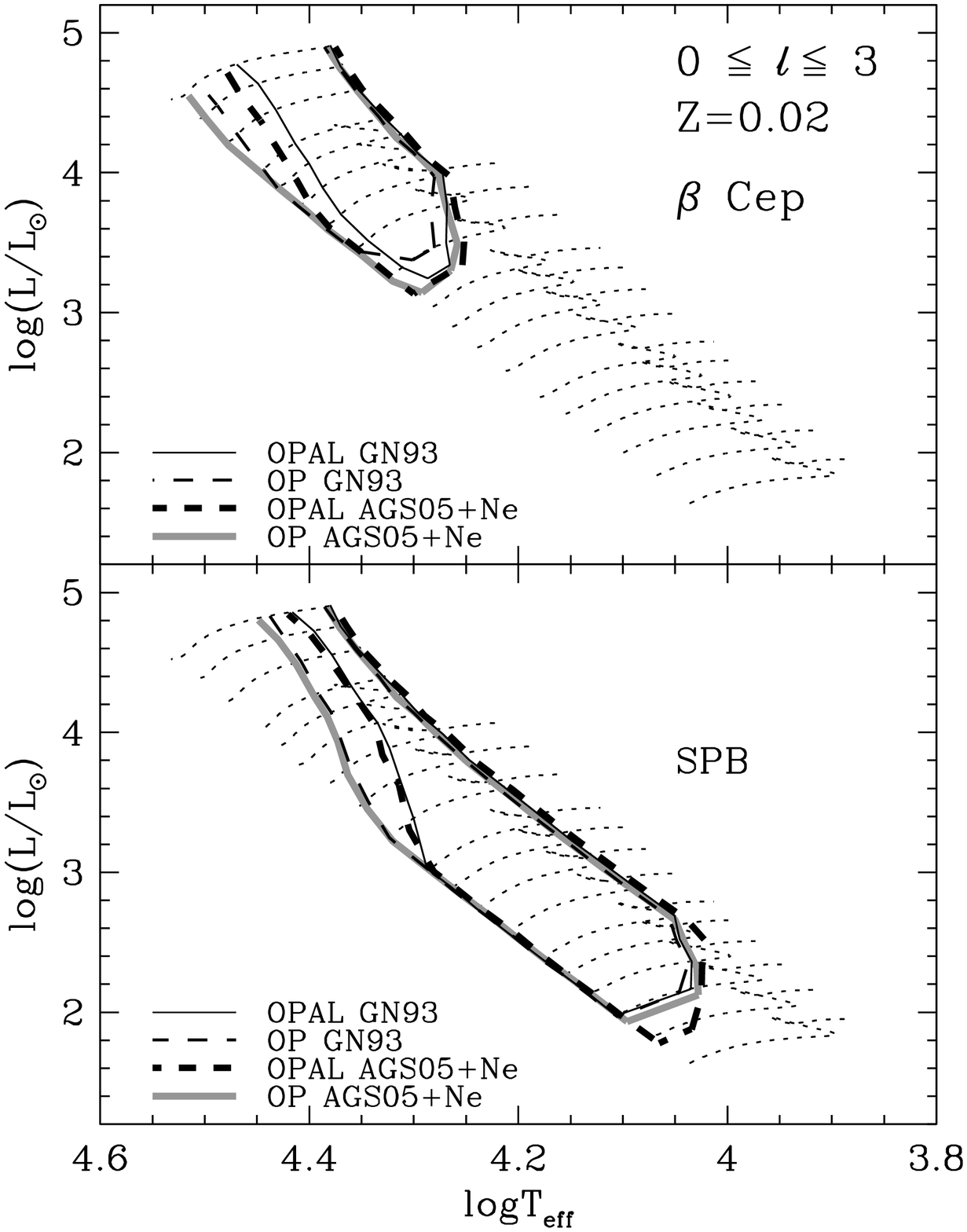}{Instability strips of $\beta$ Cep- and SPB-type pulsations in the HR diagram for Z=0.02. Evolutionary tracks are represented by dotted lines.}{fig:is_z2}{!ht}{clip,angle=0,width=.9\textwidth}
\figureDSSN{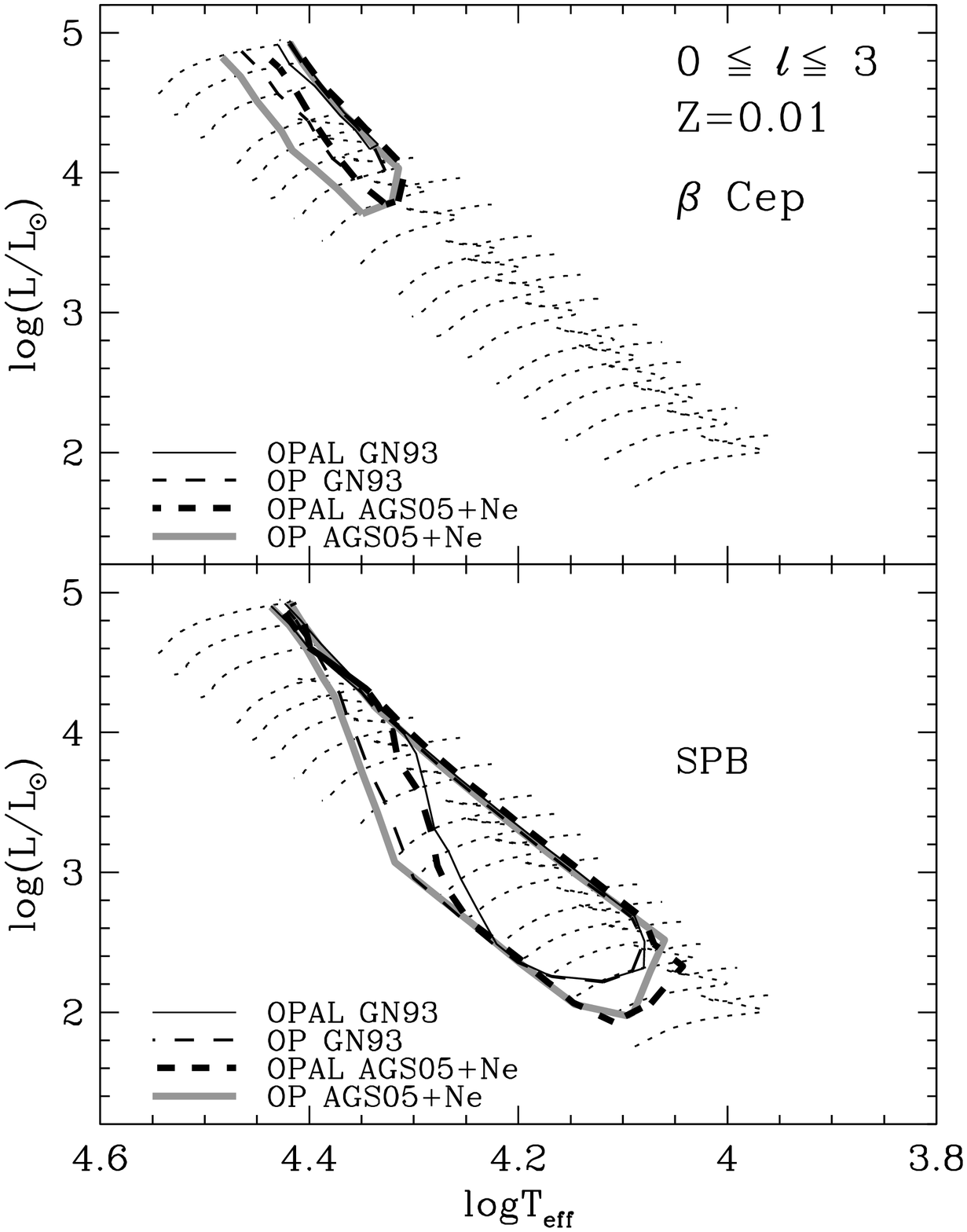}{Same as Fig. \ref{fig:is_z2} but for Z=0.01}{fig:is_z1}{!ht}{clip,angle=0,width=.9\textwidth}
\figureDSSN{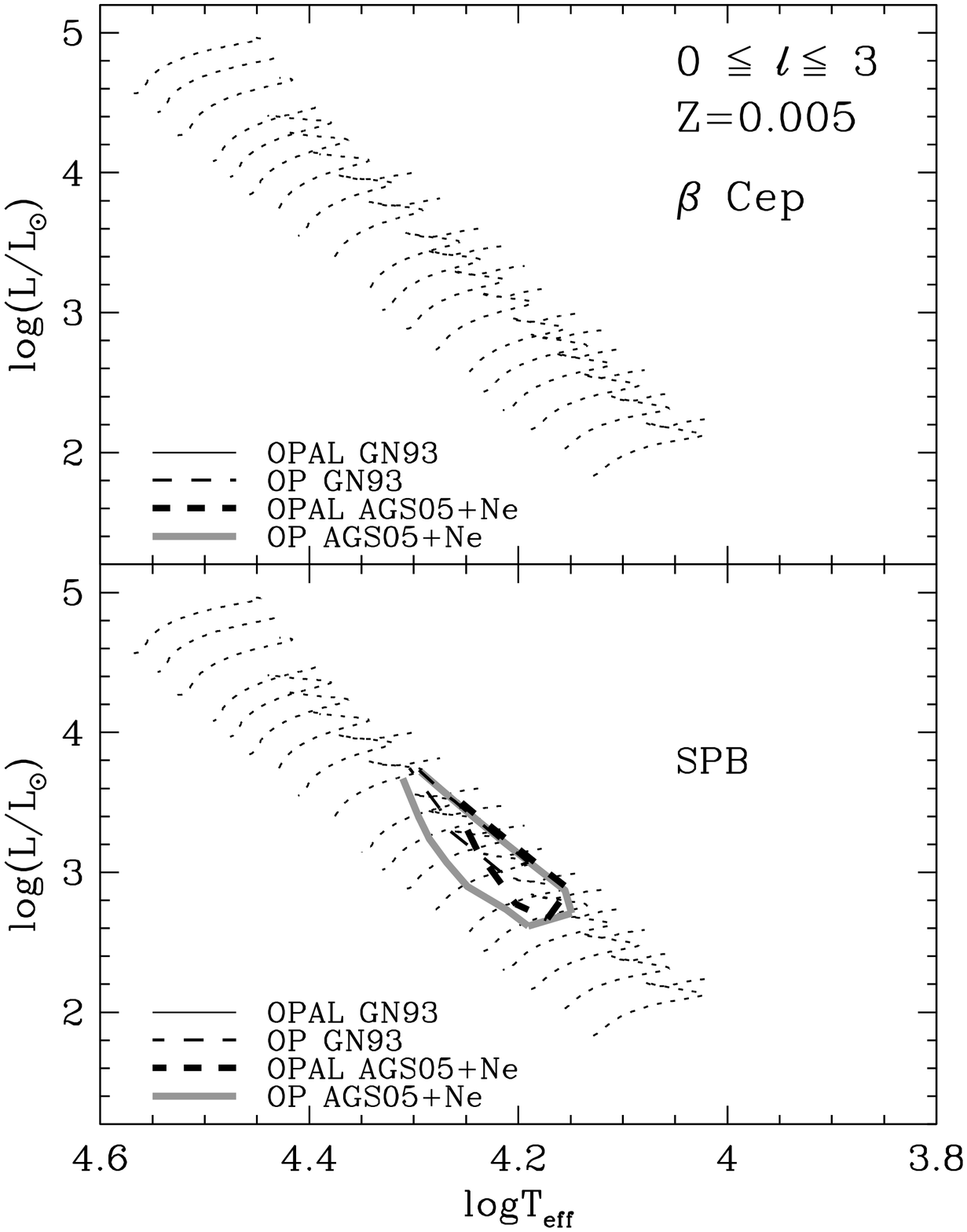}{Same as Fig. \ref{fig:is_z2} but for Z=0.005}{fig:is_z05}{!ht}{clip,angle=0,width=.9\textwidth}
\figureDSSN{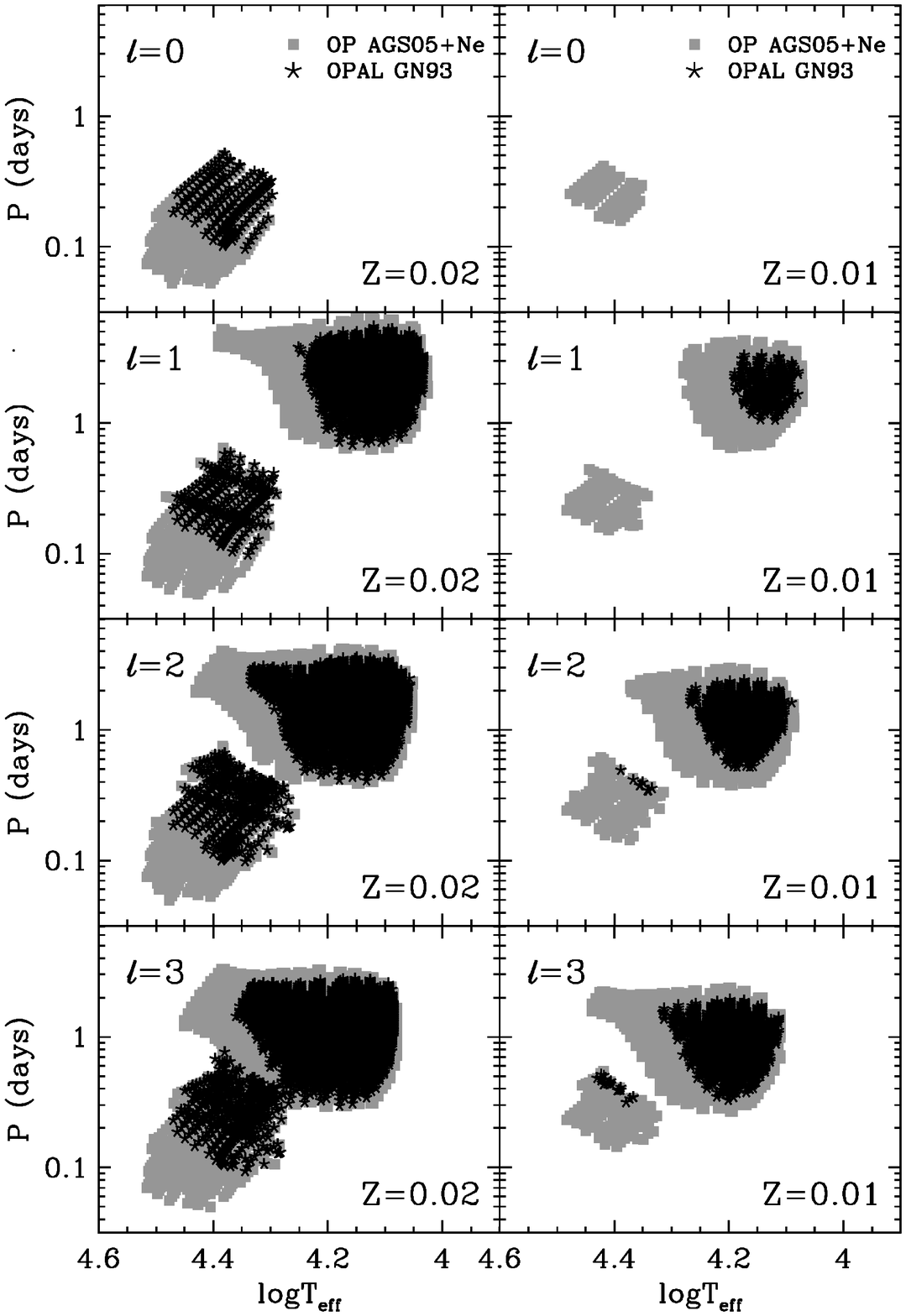}{Instability strips represented in a $\log{\teff}$-$\log{P}$ diagram for Z=0.02,0.01 and different degree $\ell$. In each panel, the two regions of unstable modes represent $\beta$ Cep- and SPB-type pulsations.}{fig:logP}{!ht}{clip,angle=0,width=0.95\textwidth}

\end{document}